\documentclass[]{elsart}
\usepackage{graphicx,amssymb,amsfonts,latexsym}     

\begin{document}

\begin{frontmatter}

\title{From single dots to interacting arrays}
\author[Reykjavik]{Vidar Gudmundsson}
\author[Reykjavik,Bucuresti]{Andrei Manolescu}
\author[Hamburg]{Roman Krahne}
\author[Hamburg]{Detlef Heitmann}
\address[Reykjavik]{Science Institute, University of Iceland,
                    Dunhaga 3, IS-107 Reykjavik, Iceland}
\address[Bucuresti]{Institutul Na\c{t}ional de Fizica Materialelor, C.P. MG-7
                    Bucure\c{s}ti-M\u{a}gurele, Rom\^ania}
\address[Hamburg]{Institut f\"ur Angewandte Physik und Zentrum
                  f\"ur Mikrostrukturforschung,\\ Universit\"at Hamburg,
                  Jungiusstra\ss e 11, D--20355 Hamburg, Germany}

%

\begin{abstract}
We explore the structural changes in charge the density  
and the electron configuration of quantum dots caused by 
the presence of other dots in an array, and the
interaction of neighboring dots. 
We discuss what recent measurements and calculation of the far-infrared  
absorption reveal about almost isolated quantum dots
and investigate some aspects of the complex transition from isolated dots
to dots with strongly overlapping electron density. We also address the
the effects on the magnetization of such dot array. 
\end{abstract}

%
%
\end{frontmatter}

\section{Introduction}
Arrays of quantum dots of different shapes and sizes have been explored
by far-infrared (FIR) absorption measurements and Raman scattering for 
a decade by many research groups. The main reason for using arrays has
been the need to increase the signal strength of the tiny quantum dots
in the weak radiation field applied, whose wavelength is up to 4 orders of   
magnitude larger than the dots. For lithographically prepared and
etched quantum dots no evidence has been found for interaction between
the dots on the length scale made available by laser holography
for periodic structures. Recently, experiments on field-effect-confined 
quantum dots in Al$_x$Ga$_{1-x}$/GaAs heterostructures have yielded
signs that have been interpreted as being caused by the periodicity 
of the confinement potential of the array \cite{Krahne01:195303}.
Evidence of direct interaction between dots in this same system have
also been found \cite{Krahne01:xx}. Here we shall review these two
cases together with the model calculations used for their interpretation.
Such inter-dot interaction had so far only been
observed for adjacent large, 20-micron-size, 2D-electron disks in
microwave experiments \cite{Dahl92:15590}.

For the parameters available in field-effect-induced arrays of quantum
dots with, typically, a lattice length of few hundred nanometers 
the interaction effects are expected to be small on the scale
of the confinement energy $\hbar\Omega$. As this energy anyways
lies in the range of few meV, where the low experimental sensitivity 
makes measurements challenging, it can be expected that mainly interaction effects 
leading to changes in the shape of the dots can be detected. With this in mind
we explore numerically the subtle effects of the interaction on the
ground state of elliptic dots in arrays with a bit shorter lattice length,
than is now common in experiments. In addition, we consider the effects
on the FIR absorption and the orbital magnetization of the dots.
The magnetoplasmon excitations in arrays of circular and noncircular
quantum dots have been studied by Zyl et al.\ in the 
Thomas-Fermi-Dirac-von Weizs{\"a}ker semiclassical approximation \cite{Zyl00:2107}. 
They study the deviations from the ideal collective excitations of isolated
parabolically confined quantum dots caused by the local perturbation
of the confining potential as well as the interdot Coulomb interaction and
find the latter indeed to be unimportant unless the interdot separation
is of the order of the size of the dots. 
An analytical model of parabolically confined electrons has been presented
with a simplified inter-dot interaction. The model predicts shifts of
collective modes and appearance of other modes that are not dipole active
\cite{Taut00:8126}.

Commonly, experimental results on the FIR absorption of quantum dots in 
Al$_x$Ga$_{1-x}$/GaAs heterostructure have successfully been interpreted
in terms of a model of an isolated quantum dot with confinement
potential that is parabolic or steeper. The pure parabolic confinement
is caused by uniformly distributed ionized donors in the
AlGaAs layer that have supplied their electrons to the active dot
layer. Furthermore, the extension of the Kohn theorem 
explains why only center-of-mass modes can be excited in such dots with
FIR radiation with wavelength much larger than the dot size 
\cite{Maksym90:108,Broido90:11400,Pfannkuche91:13132}. 
Dots satisfying these criteria thus only show a single absorption peak
at the frequency of the naked parabolic confinement. In magnetic field
the peak splits into two peaks, one approaching the cyclotron frequency
from above with increasing magnetic field strength $B$, and the other
decreasing in frequency. The two collective modes are excited with
FIR radiation with opposite circular polarization.  
In accordance with present possibilities in
sample preparation, or dot design, the most common deviations 
form the simple circular parabolic confinement studied have been; elliptic dots,
dots with weak square symmetry, and dots with quartic or stronger confinement. 
Elliptic shape of dots shows up as a simple splitting of the absorption
peak at $B=0$ \cite{Yip91:1707,Li91:5151}, 
and the square shape produces a characteristic splitting
in the upper Kohn mode at finite magnetic field 
\cite{Pfannkuche91:13132,Demel90:788a}.  
The stronger confinement
can produce a trivial blue shift and weaker absorption peaks with 
magnetic dispersion almost parallel to and above the upper Kohn mode
\cite{Gudmundsson91:12098,Ye94:17217}. 
In addition, in confinement potentials that do not fullfil the criteria
for the Kohn theorem so called Bernstein modes are excited causing
characteristic splitting in the upper Kohn mode around higher
harmonics of the cyclotron frequency \cite{Gudmundsson95:17744,Bernstein58:10}.     
Interestingly enough, researchers have been able to produce ring shaped
quantum dots and measure their FIR absorption, but these do not form regular arrays
\cite{Lorke00:2223,Zaremba96:R10512,Llorens01:035309,Emperador00:4573}.

\section{Effects of an array}
\begin{figure}
\begin{center}
  \includegraphics[bb=90 433 300 694,clip,width=12cm]{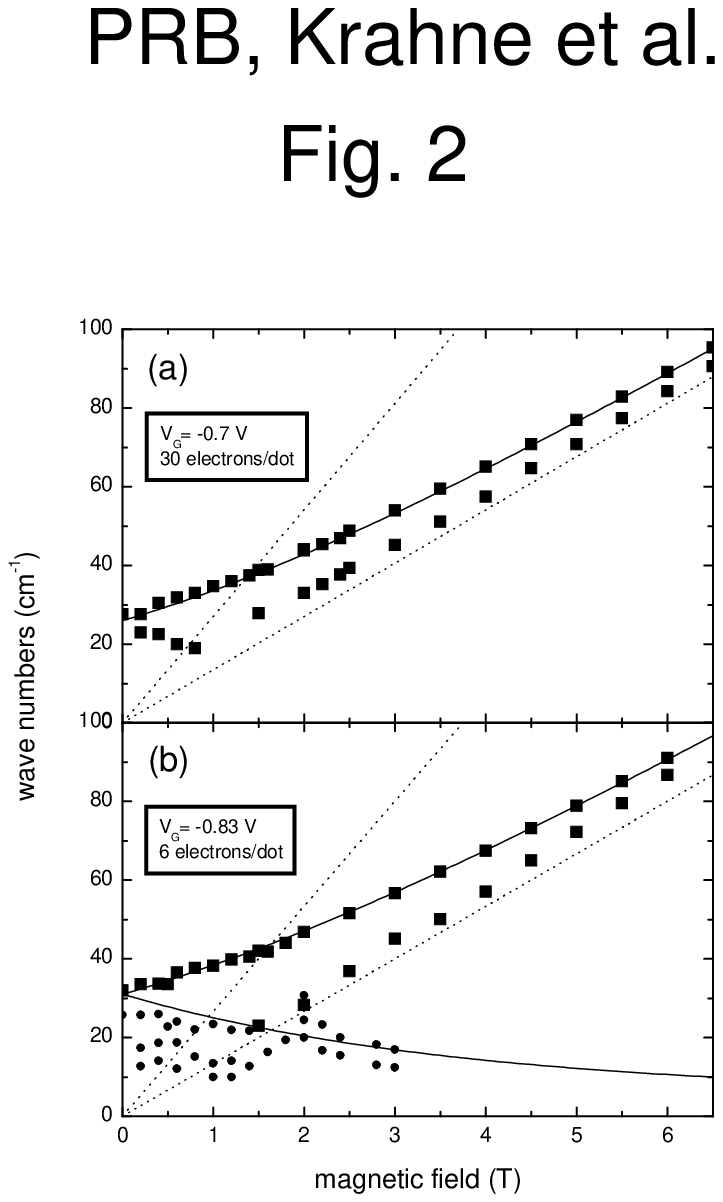}
\end{center}
\caption{Experimental dispersion for quantum dots with (a) 30
         electrons and (b) 6 electrons. Full lines are fits with the Kohn
         modes of eq.(1), the dotted lines are $\omega_c$ and 2$\omega_c$
         extracted from this fit. A new mode, the below-Kohn mode, is
         observed {\it below} the upper Kohn mode but clearly {\it above}
         $\omega_c$.}
\label{Fig_RK_u2om}
\end{figure}

The simplest effects of an array of quantum dots on the confinement potential
of an individual dot would be the eventual flattening of the potential imposed
by the periodicity of the array. In field-induced dot arrays, where each dot
contains only few electrons,
it must be possible to have the confinement potential shallow enough that at least
electrons in the excited states are affected by this weakening of the
confinement. This has been demonstrated by Krahne et al.\ \cite{Krahne01:195303}.
The experimental dispersion curves are seen in Fig.\ \ref{Fig_RK_u2om} for
6 or 30 electrons in each quantum dot. A purely parabolically confined quantum
dot has the FIR dispersion of the Kohn modes 
\begin{equation}
      \omega_{\pm}=\sqrt{\Omega^2_0+(\omega_c/2)^2}\pm\omega_c/2,  
\end{equation}
where $\Omega_0$ is the confinement frequency, and $\omega_c=eB/(m^*c)$ is the
cyclotron frequency. We have fitted this dispersion with the lower absorption
branch and the sharper upper one in the experimental dispersion 
curves displayed in Fig.\ \ref{Fig_RK_u2om}.
In addition to these two branches the experiments show a third
branch just below $\omega_+$ however, above the cyclotron frequency $\omega_c$.

\begin{figure}
\begin{center}
  \includegraphics[width=12cm]{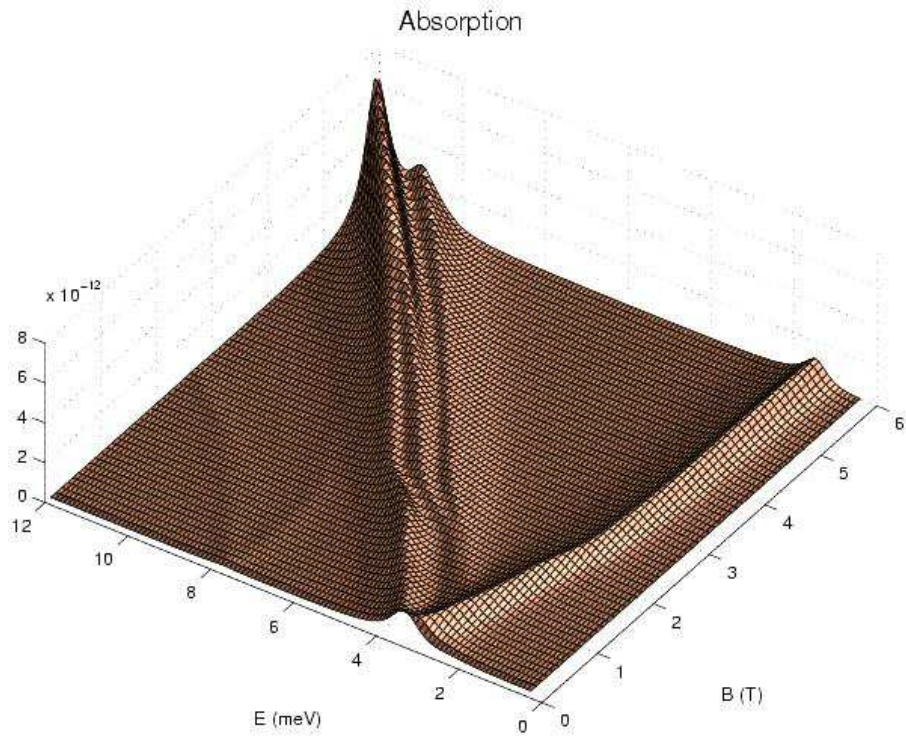}
\end{center}
\caption{Calculated dipole absorption for a quantum dot with 5 electrons
         in a flattened potential described in the text.
         In addition to the strong Kohn modes new
         modes below the high-frequency Kohn mode are found also in the
         calculation. The half-linewidth is 0.3 meV and $T=1$ K.}
\label{Fig_VG_3Disog}
\end{figure}

It is well known that the energy spectrum of electrons in a periodic lattice
can be calculated only for a a magnetic flux commensurable with the unit cell
\cite{Hofstadter76:2239}.
It is technically difficult to vary the magnetic field continuously 
to describe the experimental results for an array of dots with interacting electrons 
\cite{Gudmundsson95:16744,Gudmundsson96:5223R}. We thus resort to a model of
an individual quantum dot in the Hartree approximation, but with a potential
of the form
\begin{equation}
      V(x)=ax^2+bx^4+W(x), 
\end{equation}
where $x=r/a_0^*$ is the radial coordinate scaled by the
effective Bohr radius $a_0^*=9.77$ nm in GaAs and
\begin{equation}
      W(x)=c\left[1-f\left( 3.9x-12 \right)\right],
\end{equation}
with $f(x)=1/(\exp(x)+1)$. The calculated FIR absorption is shown in 
Fig.\ \ref{Fig_VG_3Disog} for the parameter choice $a=0.48$ meV, 
$b=-1.8\time 10^{-3}$ meV, and $c=6$ meV. These parameters have been  
selected to give results qualitatively close to the experiment, 
without performing an actual fit. The model yields a mode 
just below $\omega_+$ as is seen in the experiment. At low magnetic field
the upper Kohn branch, $\omega_+$, has a complex splitting around
$\omega =2\omega_c$ that is dependent on the number of electrons
present and for a higher number of electrons develops into a 
splitting caused by a Bernstein mode \cite{Gudmundsson95:17744,Bernstein58:10}.
At high magnetic field the induced density of the $\omega_+$ mode indeed
reflects a center-of-mass mode, but the lower mode, the new one, is the
lowest internal mode with one node in the center of the 
dot \cite{Krahne01:195303}. 
For a confinement stronger than the parabolic one 
(for example, with $b>0$ and $c=0$) this internal mode is
usually found above the upper Kohn mode, but here due to the special
confinement it has lower energy. 

This is clearly an effect of the shape of the confinement potential of an
individual quantum dot in an array, but what about a direct interaction between
dots?

\section{Interaction between dots}
\subsection{Experimental indications}
\begin{figure}
\begin{center}
  \includegraphics[hiresbb,clip,width=12cm]{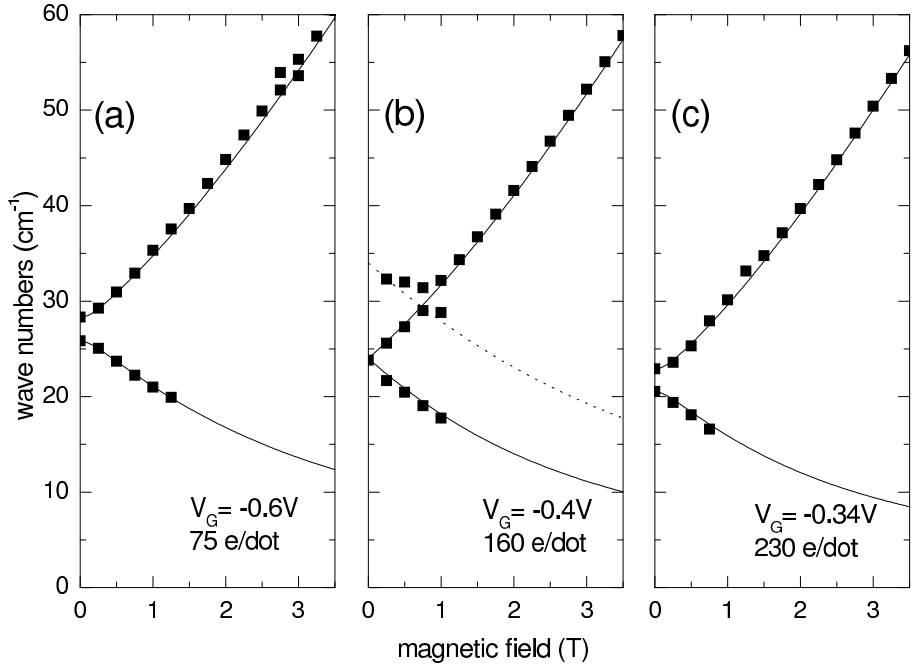} 
\end{center}
\caption{Magnetic field dispersions for three different gate
         voltages. The experimental resonance positions extracted from the
         spectra are depicted by the full squares. The solid lines show a
         calculation according to eq.\ (\ref{EDdisp}) with $\omega_{x(y)}$ as
         fit parameters. (a) $V_G=-0.6$ V: strong confinement leading to
         isolated dots. (b) $V_G=-0.4$ V: weaker confinement. Here an
         anticrossing of the $\omega_+$ mode around $B=1$ T with another
         weak resonance, which decreases in energy with increasing $B$, is
         observed. This anticrossing is behavior is a characteristic
         property of square symmetric quantum dots \cite{Demel90:788a}. (c)
         $V_G=-0.34$ V: weak confinement leading to overlapping dots.}
\label{Fig_RK_elsqDisp}
\end{figure}

Indications for interaction between quantum dots have been found 
in the same type of system when the dots have been prepared to have
an elliptic symmetry rather than the circular one \cite{Krahne01:xx}.
In elliptical quantum dots the rotational symmetry is broken and the
degeneracy of the $\omega_+$ and $\omega_-$ modes is lifted at
$B=0$. The dispersion of the FIR absorption peaks in a single elliptic 
quantum dot with parabolic confinement is described by
\begin{equation}
      \omega_\pm^2 = \frac{1}{2} \left( \omega_x^2 + \omega_y^2 +
      \omega_c^2 \pm \sqrt{\omega_c^4 + 2 \omega_c^2 ( \omega_x^2 +
      \omega_y^2 ) + (\omega_x^2 - \omega_y^2)^2} \:\right),
\label{EDdisp}
\end{equation}     
where $\omega_x$ and $\omega_y$ are, respectively, the resonance frequencies for the
oscillation in the $x$ and $y$ direction at $B=0$, the two symmetry axis of the
dot. Let us consider the long axis of the ellipse to be aligned 
with the $y$ axis of the coordinate system.
The two modes $\omega_+(B=0)=\omega_x$ and $\omega_-(B=0)=\omega_y$
are observed with orthogonal linear polarization \cite{Krahne01:xx}. 
Figure \ref{Fig_RK_elsqDisp}
shows the magnetic dispersion of the absorption peaks for 3 different
values of the gate voltages which is used to 
control the number of electrons in each quantum dot. 
For few electrons, Fig.\ \ref{Fig_RK_elsqDisp}(a), the dispersion 
with two peaks at $B=0$ reflects the elliptic shape of the quantum dots. 
Curiously enough, for a higher number of electrons, Fig.\ \ref{Fig_RK_elsqDisp}(b),
the peaks at $B=0$ are degenerate and the curves conform with the dispersion
measured \cite{Demel90:788a} and calculated \cite{Pfannkuche91:13132} 
for square shape dots with a characteristic anticrossing in the
$\omega_+$ mode at a nonvanishing magnetic field. At even higher electron number,
Fig.\ \ref{Fig_RK_elsqDisp}(c), the characteristic traits of the square
shape are lost, no anticrossing at finite $B$ and no degeneracy of the
modes at $B=0$ is discernible, but now the dispersion can be well fitted
by the formula for elliptic dots (\ref{EDdisp}) again.
Linearly polarized measurements show that for all gate voltages
the energetically higher excitation is polarized in $x$ direction
and the energetically lower in
$y$ direction. This shows that no rotation of the ellipse out of
some electrostatic reason takes place. Thus the actual geometrical
shape of the dots must be deformed by some interaction with their
neighbors in dependence of the gate voltage.

\subsection{Model results for interacting dots; Ground state properties}  
\begin{figure}
\begin{center}
  \includegraphics[width=12cm]{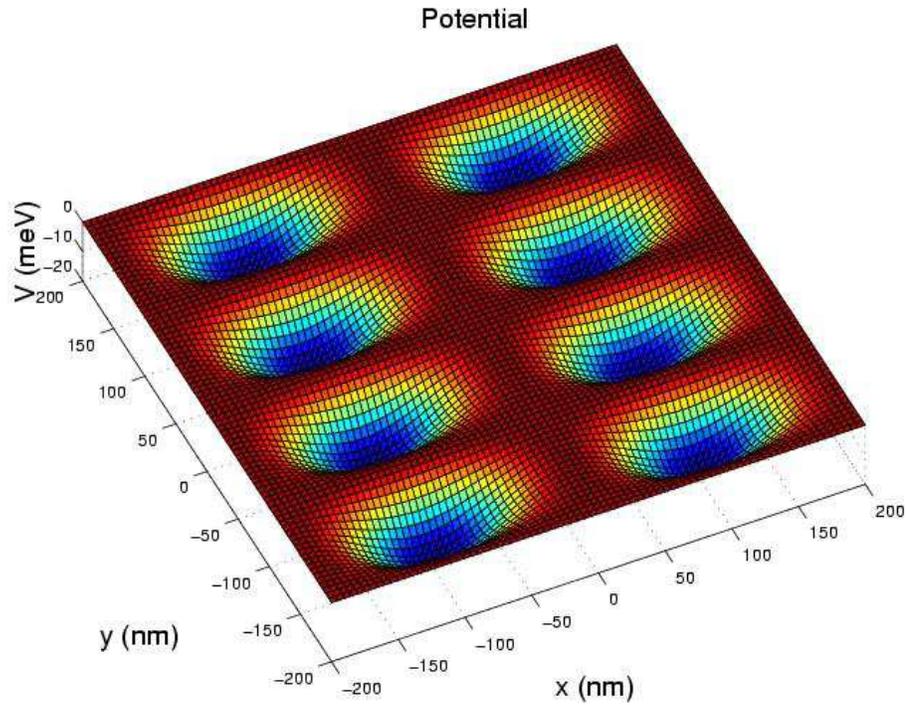}
\end{center}
\caption{The periodic confinement potential $V_{\mbox{\scriptsize QAD}}$ 
         for a dot array with
         aspect ratio 1:2. See equation (\ref{V_QAD}) in text.}
\label{Vm_pq8}
\end{figure}

\begin{figure}
\begin{center}
  \includegraphics[width=12cm]{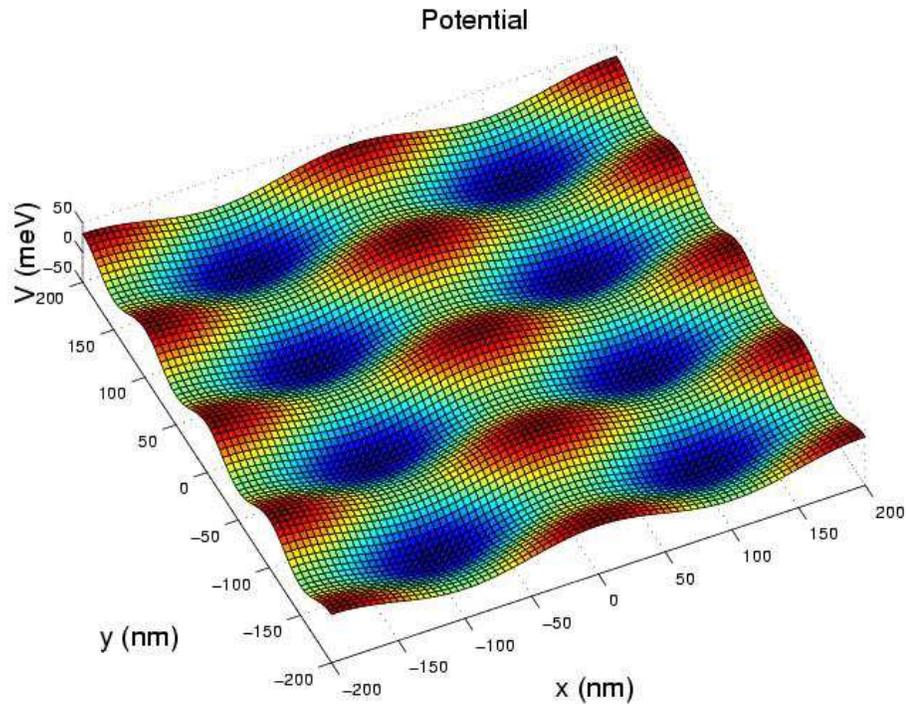}
\end{center}
\caption{The simple cosine confinement potential $V_{\mbox{\scriptsize per}}$ 
         with aspect ratio 1:2. See equation (\ref{V_per}) in text.}
\label{Vcos_pq8}
\end{figure}

We model an array of quantum dots as interacting electrons in a periodic 
potential. We choose to describe the interaction of electrons within a dot
at the same level as the interaction of electrons in different dots.
In order to distinguish between the effects of different parts of the
interaction, the direct one, the exchange,  and correlation part, 
we use density functional theory (DFT) approach in the 
local spin density approximation (LSDA) to be described below. 
We describe a simple array of circular or elliptic dots (or antidots) 
shaped by the potential
\begin{equation}
      V_{\mbox{\scriptsize QAD}}({\bf r})=V_0\left[\sin\left(\frac{g_1 x}{2}\right)
                 \sin\left(\frac{g_2 y}{2}\right)\right]^2,
\label{V_QAD}    
\end{equation}
where $g_i$ is the length of the fundamental inverse lattice vectors, ${\bf
g}_1=2\pi \hat{\bf x}/L_x$, and ${\bf g}_2=2\pi \hat{\bf y}/L_y$. The Bravais
lattice defined by $V_{\mbox{\scriptsize QAD}}$ has period lengths
$L_x$, $L_y$, and the inverse lattice is spanned by ${\bf G}=G_1{\bf g}_1+G_2{\bf
g}_2$, with $G_1,G_2\in {\bf Z}$. The commensurability condition between the
magnetic length $\ell=(\hbar c/(eB))^{1/2}$ and the periods $L_i$ 
requires magnetic-field values
of the form $B=pq\phi_0/L_xL_y$, with $pq\in {\bf N}$ flux quanta, $\phi_0=hc/e$,
in a unit cell \cite{Silberbauer92:7355,Gudmundsson95:16744}.
Arbitrary rational values can, in principle, be obtained by resizing
the unit cell in the Bravais lattice, but numerically this is quite difficult.
The term 'circular quantum dot' can of course only describe a dot with few
electrons in a square lattice where the electron density is concentrated in
the middle of the cell and vanishes well before the cell edge. 
As the number of electrons increases the
electron density must reflect the symmetry of the lattice. The dot
potential (\ref{V_QAD}) is seen in Fig.\ \ref{Vm_pq8}. 
We investigate different confinement potentials later on in order to
understand better the interaction between dots. One of interest
will be the simple periodic cosine potential 
\begin{equation}
      V_{\mbox{\scriptsize per}}({\bf r})=V_0\cos( g_1x)+V_0\cos(g_2y),
\label{V_per}
\end{equation}
shown in Fig.\ \ref{Vcos_pq8} for an array of elliptic dots.

The exchange-correlation energy per particle, $\epsilon_{xc}(\nu ,\xi )$,
is parameterized in terms of the total filling factor 
$\nu =\nu_{\uparrow}+\nu_{\downarrow}=2\pi\ell^2n$ and the spin polarization
\begin{equation}
      \xi = \frac{\nu_{\uparrow}-\nu_{\downarrow}}
            {\nu_{\uparrow}+\nu_{\downarrow}}
\end{equation}
rather than the spin densities $n_{\uparrow}$ and $n_{\downarrow}$ 
\cite{Lubin97:10373}. The exchange-correlation potentials are then
\begin{equation}
      V_{xc{\uparrow\atop\downarrow}}=\frac{\partial}{\partial\nu}
      (\nu\epsilon_{xc})\pm (1\mp\xi )\frac{\partial}{\partial\xi }
      \epsilon_{xc},      
\end{equation}
and the exchange-correlation energy is interpolated as 
\begin{equation}
      \epsilon_{xc}(\nu ,\xi )=\epsilon^{\infty}_{xc}(\nu )e^{-f(\nu )}+
      \epsilon_{xc}^{\mbox{\scriptsize TC}}(1-e^{(1-f(\nu )})\quad\mbox{with}\quad
      f(\nu )=1.5\nu+7\nu^4
\end{equation}
between the infinite magnetic
field limit $\epsilon^{\infty}_{xc}=-0.782\sqrt{2\pi n}e^2/\kappa$ and the zero
field limit $\epsilon^{\mbox{\scriptsize TC}}_{xc}$ given by Tanatar and Ceperly
\cite{Tanatar89:5005} generalized to intermediate polarizations 
\cite{Koskinen97:1389}. 

In the numerical calculations we shall assume the magnetic flux density 
through the relevant unit cell to be constant and set to $B=1.654$ T
leading to a magnetic length $\ell =19.95$ nm much smaller than the
lattice lengths $L_x$ and $L_y$, which shall be in the range 100 to
200 nm. It should be emphasized here that since the Coulomb interaction
is treated equally for all electrons in the system, independent of 
whether they are in the same dot or not, we turn it totally off when
we discuss a noninteracting system.   

\begin{figure}
\begin{center}
  \includegraphics[angle=90,width=12cm]{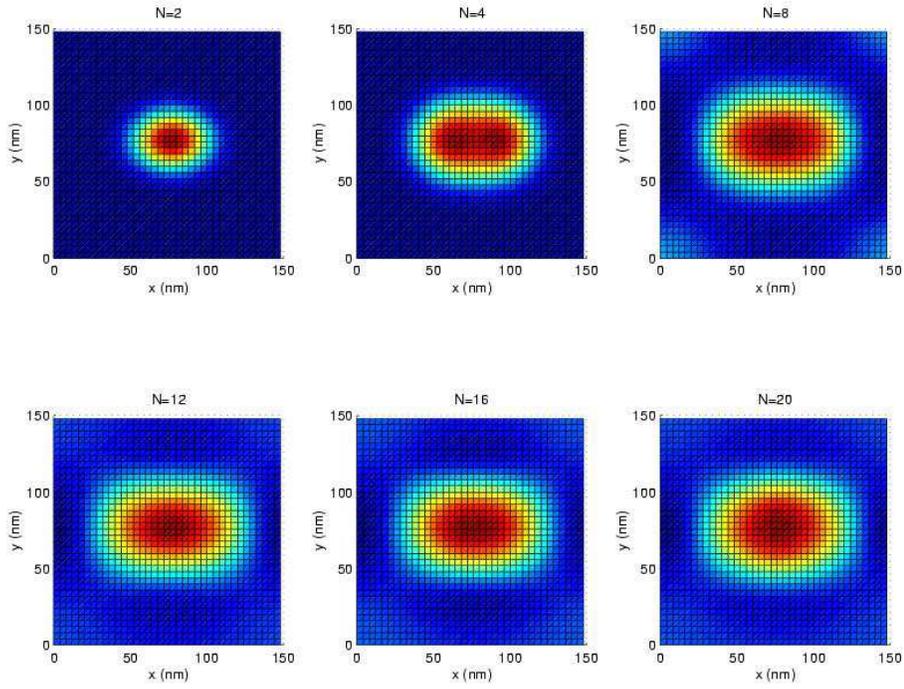}
\end{center}
\caption{The electron density distribution for the ground state of
         interacting elliptical dots in a square lattice. The confinement
         is according to eq.\ (\ref{V_sq}). $B=1.654$ T, $T=1$ K, $V_0=-16$ meV.}
\label{Fig_VG_DeRK}
\end{figure}

The resulting Kohn-Sham equations are solved within the symmetric Ferrari
basis \cite{Ferrari90:4598,Silberbauer92:7355,Gudmundsson95:16744} and
in order to have a large enough basis allowing several electrons in each 
dot we have to use lattice lengths $L_i$ shorter than the actual ones 
in experiments. In the present calculations we use
upto 16384 basis states. To get back to the experimental results displayed in
Fig.\ \ref{Fig_RK_elsqDisp} we perform a calculation for the ground
state properties of an array of dots described by the confinement
potential    
\begin{equation}
      V_{\mbox{\scriptsize sq}}({\bf r})=V_0\left[\sin\left(\frac{g_1 x}{2}\right)
                 \left\{\sin\left(\frac{g_2 y}{2}\right)\right\}^2\right]^2.
\label{V_sq}
\end{equation}
This choice defines a square unit cell, but within each cell the
dot confinement is elliptic. The model results are shown as density contours in 
Fig.\ \ref{Fig_VG_DeRK} for a growing number of electrons. For few electrons
the shape of the dots is very close to circular, but with increasing electron
number the dots become more elliptic. For still a higher number of electrons
the Coulomb repulsion between the narrower edges (their ends) of neighboring  
dots causes their shape
to become more square like. For $N=20$, when their density clearly overlaps
the central region acquires a circular or elliptic form again. Now, we can
not maintain that this demonstrates what happens in an absorption experiment, but 
by a comparison to the noninteracting case we can clearly see that the Coulomb 
interaction has a strong influence on the the shape of the dots as the number
of electrons is increased.  

\begin{figure}
\begin{center}
  \includegraphics[width=12cm]{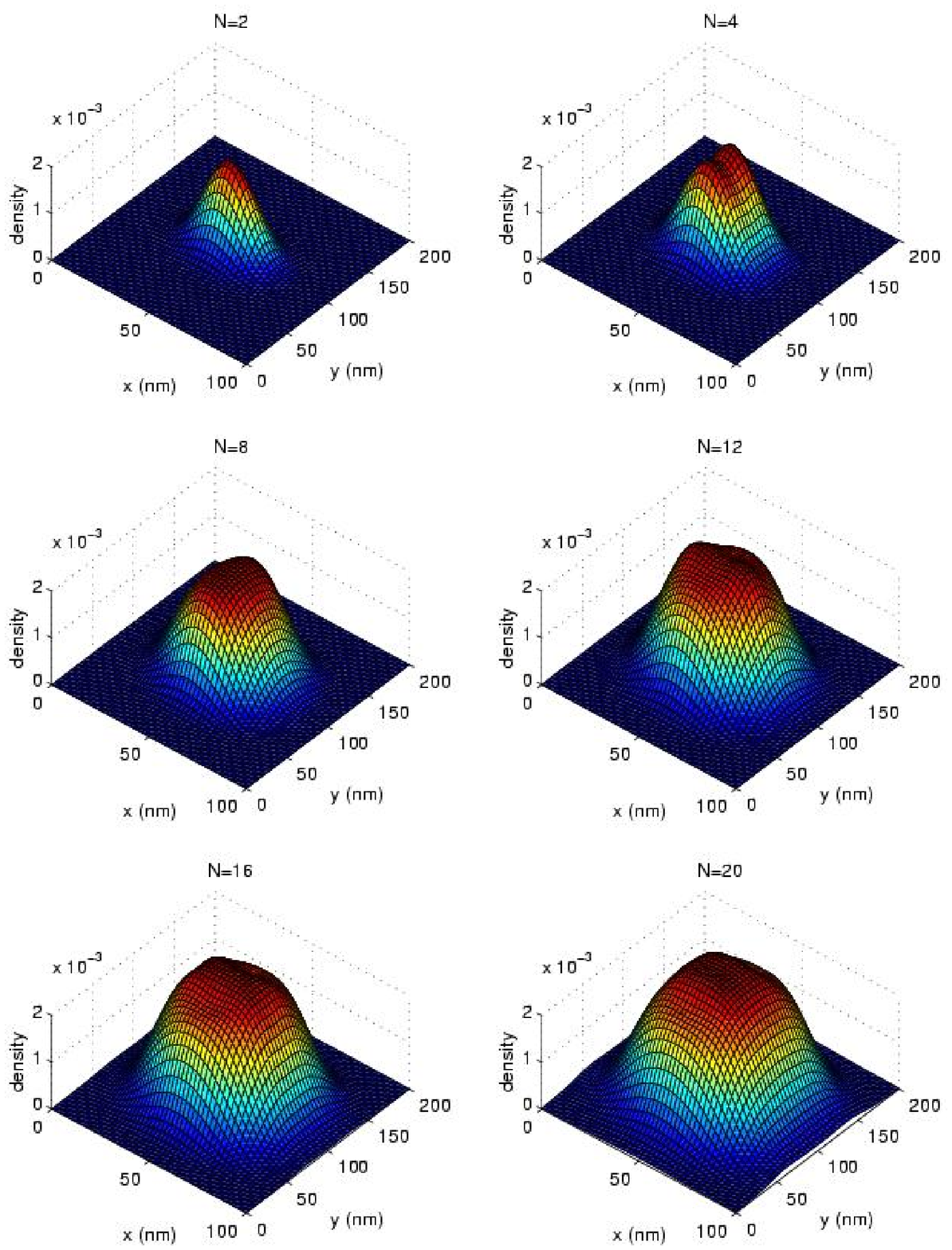}
\end{center}
\caption{The electron density distribution for the ground state of
         {\it noninteracting} elliptical dots. 
         The confinement is according to eq.\ (\ref{V_QAD}).
         The $x$ and $y$ axis are scaled differently here.
         $B=1.654$ T, $T=1$ K, $V_0=-16$ meV.}
\label{RoDe20_m}
\end{figure}

\begin{figure}
\begin{center}
  \includegraphics[width=12cm]{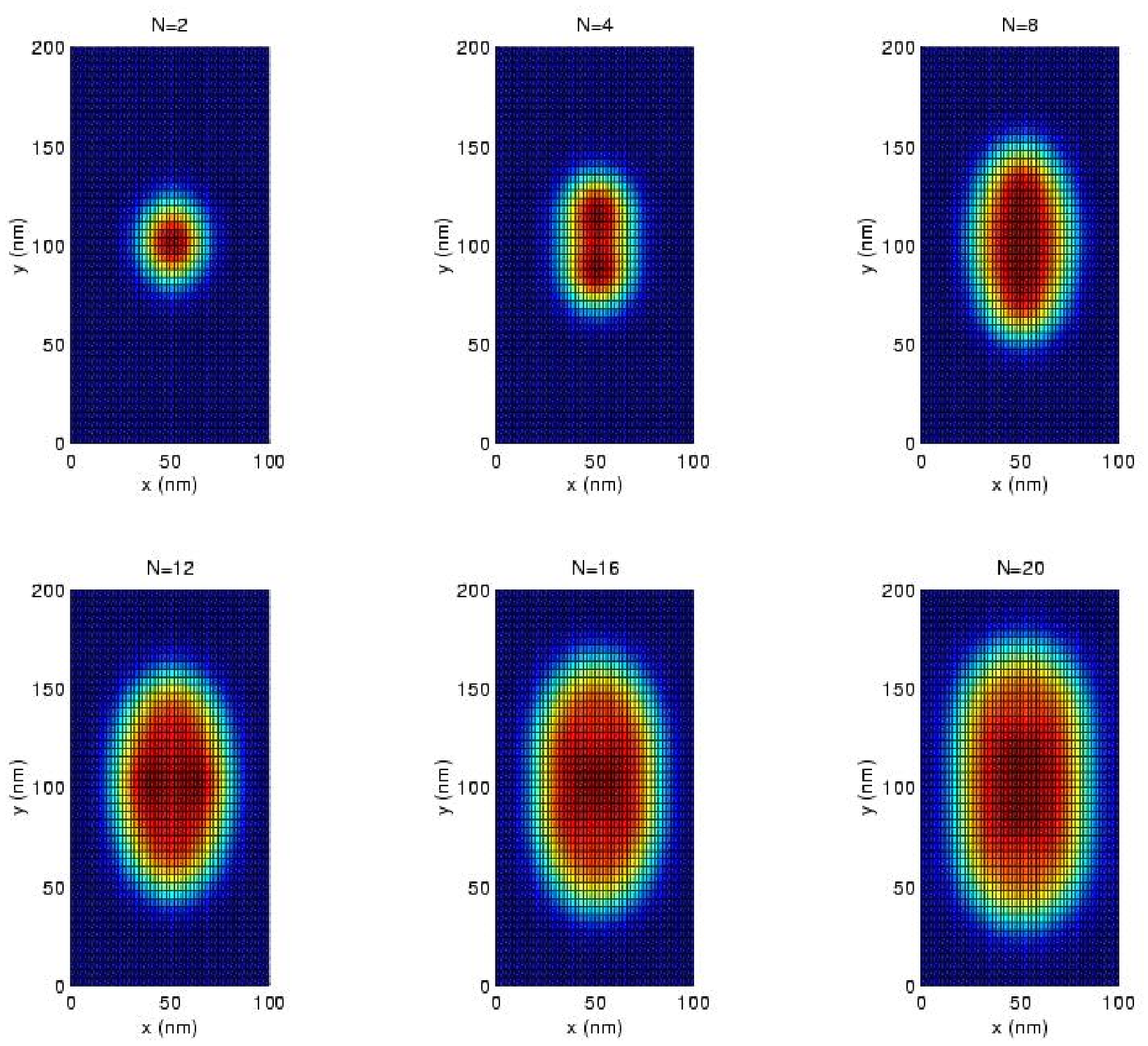}
\end{center}
\caption{The electron density distribution for the ground state of
         {\it noninteracting} elliptical dots.
         The confinement is according to eq.\ (\ref{V_QAD}).
         Same case as in Fig.\ \ref{RoDe20_m}. $B=1.654$ T, $T=1$ K,
          $V_0=-16$ meV.}
\label{RoDe20_Zm}
\end{figure}

To learn more about these effects we want to compare the results to what
happens in the different confinement potentials 
(\ref{V_QAD}) and (\ref{V_per}) introduced above.
We start with elliptic dots described by $V_{\mbox{\scriptsize QAD}}$ 
in eq.\ (\ref{V_QAD})in a rectangular lattice. Although the lattice does
not have a square shaped unit cell as the one defined by 
$V_{\mbox{\scriptsize sq}}$, the distance between the quantum dots 
measured in the $x$ or the $y$
direction can be expected to be comparable. The density for the noninteracting
case can be seen in Fig.\ \ref{RoDe20_m} with the contours displayed 
in Fig.\ \ref{RoDe20_Zm} with the real aspect ratio between the $x$ and the
$y$ axis. Up to $N=20$ there is only a weak overlapping of the electron
density of neighboring cells, but certainly the shape of the dots changes
with growing $N$, even in the absence of the electron-electron interaction.
For low $N$ their ellipticity increases as $N$ grows, but for higher 
electron number the shape slowly approaches the symmetry of the lattice.

\begin{figure}
\begin{center}
  \includegraphics[width=12cm]{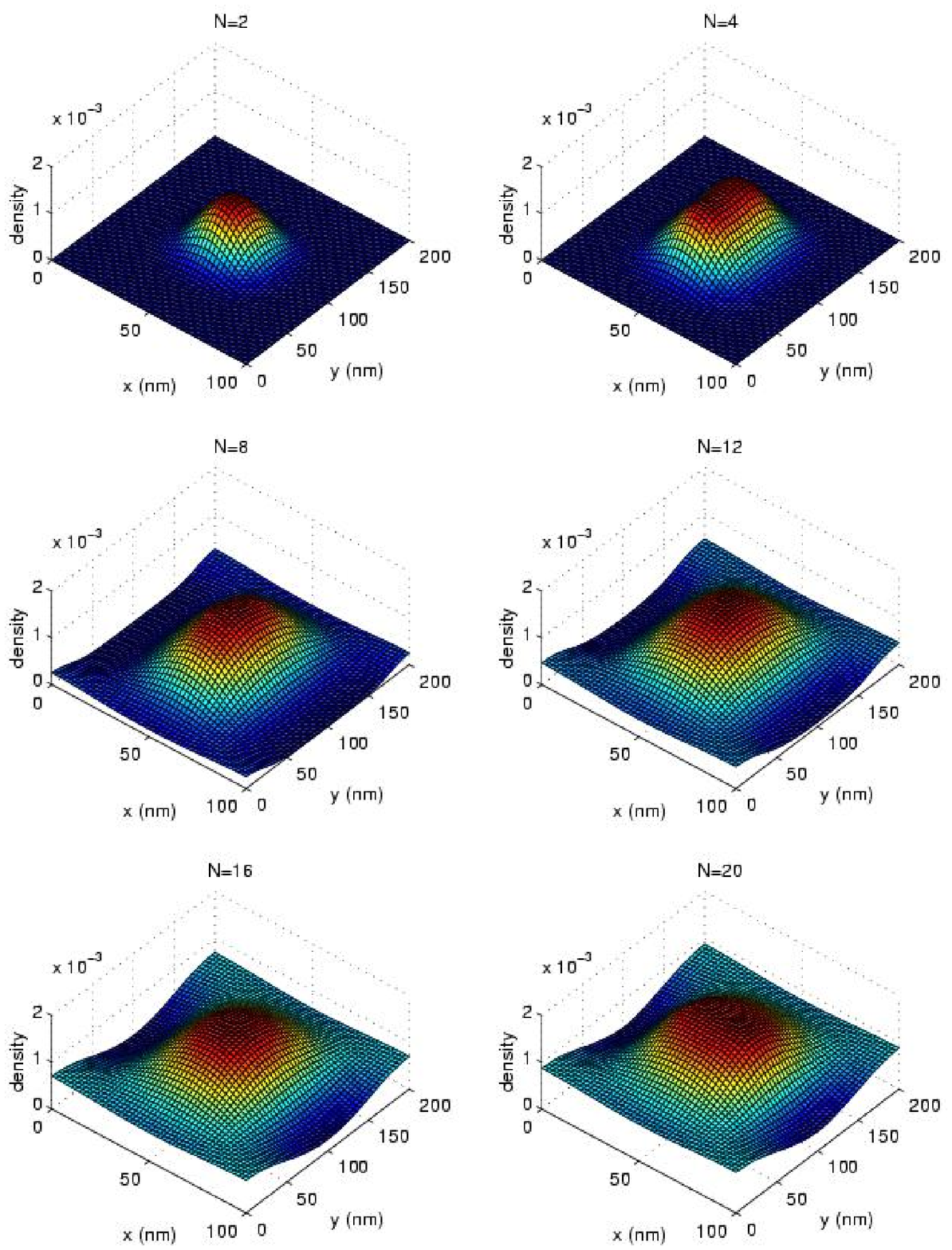}
\end{center}
\caption{The electron density distribution for the ground state of
         {\it interacting} elliptical dots.
         The confinement is according to eq.\ (\ref{V_QAD}).
         The $x$ and $y$ axis are scaled differently here.
         $B=1.654$ T, $T=1$ K, $V_0=-16$ meV.}
\label{RoDe2lda_m}
\end{figure}

\begin{figure}
\begin{center}
  \includegraphics[width=12cm]{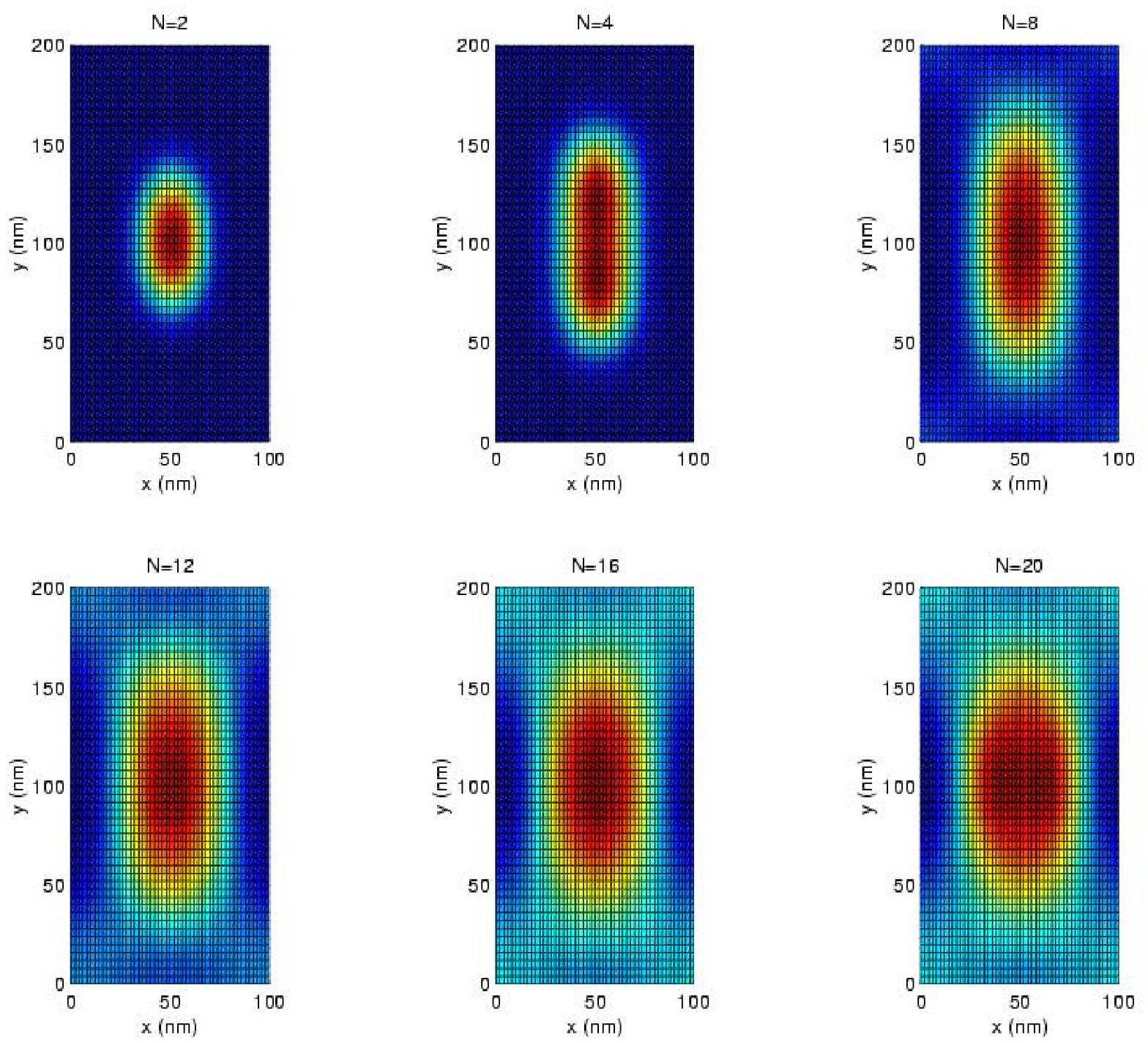}
\end{center}
\caption{The electron density distribution for the ground state of
         {\it noninteracting} elliptical dots.
         The confinement is according to eq.\ (\ref{V_QAD}).
         Same case as in Fig.\ \ref{RoDe2lda_m}. $B=1.654$ T, $T=1$ K,
         $V_0=-16$ meV.}
\label{RoDe2lda_Zm}
\end{figure}

\begin{figure}
\begin{center}
  \includegraphics[width=12cm]{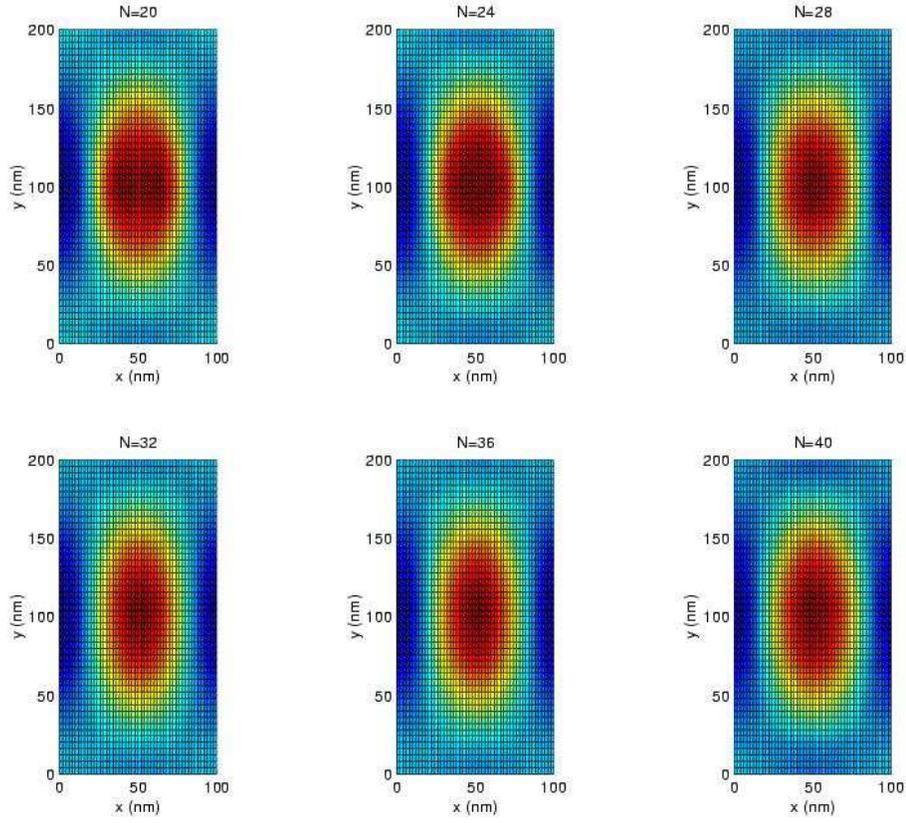}
\end{center}
\caption{The electron density distribution for the ground state of
         {\it noninteracting} elliptical dots.
         The confinement is according to eq.\ (\ref{V_QAD}).
         Same system as in Fig.\ \ref{RoDe2lda_Zm}, but with a higher
         number of electrons. $B=1.654$ T, $T=1$ K,  $V_0=-16$ meV.}
\label{RoDe2lda_XZm}
\end{figure}

The electron density for the interacting case can be seen in 
Fig.\ \ref{RoDe2lda_m} together with the contours in Fig.\ \ref{RoDe2lda_Zm}.
In contrast to the noninteracting case the Coulomb repulsion is strong 
enough to push the electron density to already overlap considerably
at $N=12$ or even a bit lower. Interestingly, the repulsion forces
the electrons to form wires in the direction of the longer axis ($y$ axis) of the
ellipses. This behavior is continued to much higher number of electrons as 
can be verified in Fig.\ \ref{RoDe2lda_XZm}, and it can be understood 
as governed by two facts. First, the repulsion between the dots along
the longer edge is stronger than between the shorter edges of the dots.
Second, the lower slope of the confinement potential in the $y$ direction
than in the $x$ direction determines an asymmetric screening in
the electron system.
In this connection, it is also clear that the electronic density in 
the elliptic dots in the square lattice defined by 
$V_{\mbox{\scriptsize sq}}$ had more space to 'broaden' the dots by
occupying the space along the long edge between them than in this system. 

\begin{figure}
\begin{center}
  \includegraphics[width=12cm]{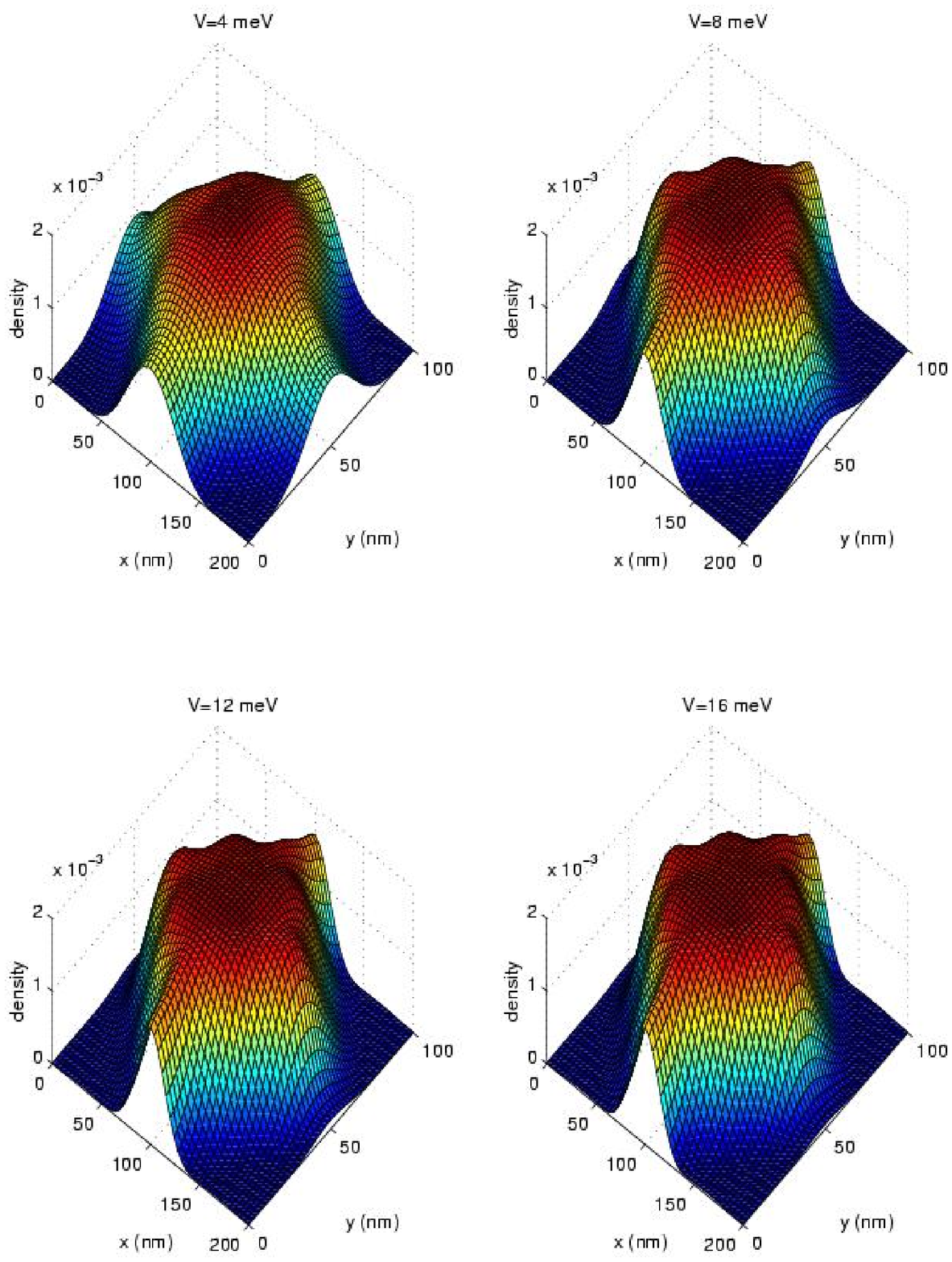}
\end{center}
\caption{The electron density distribution for the ground state of
         a {\it noninteracting} 2DEG in the simple periodic potential
         $V_{\mbox{\scriptsize per}}$ described by eq.\ (\ref{V_QAD}).
         The $x$ and $y$ axis are scaled differently here.
         $B=1.654$ T, $T=1$ K, $N=20$.}
\label{RoV0}
\end{figure}

\begin{figure}
\begin{center}
  \includegraphics[width=12cm]{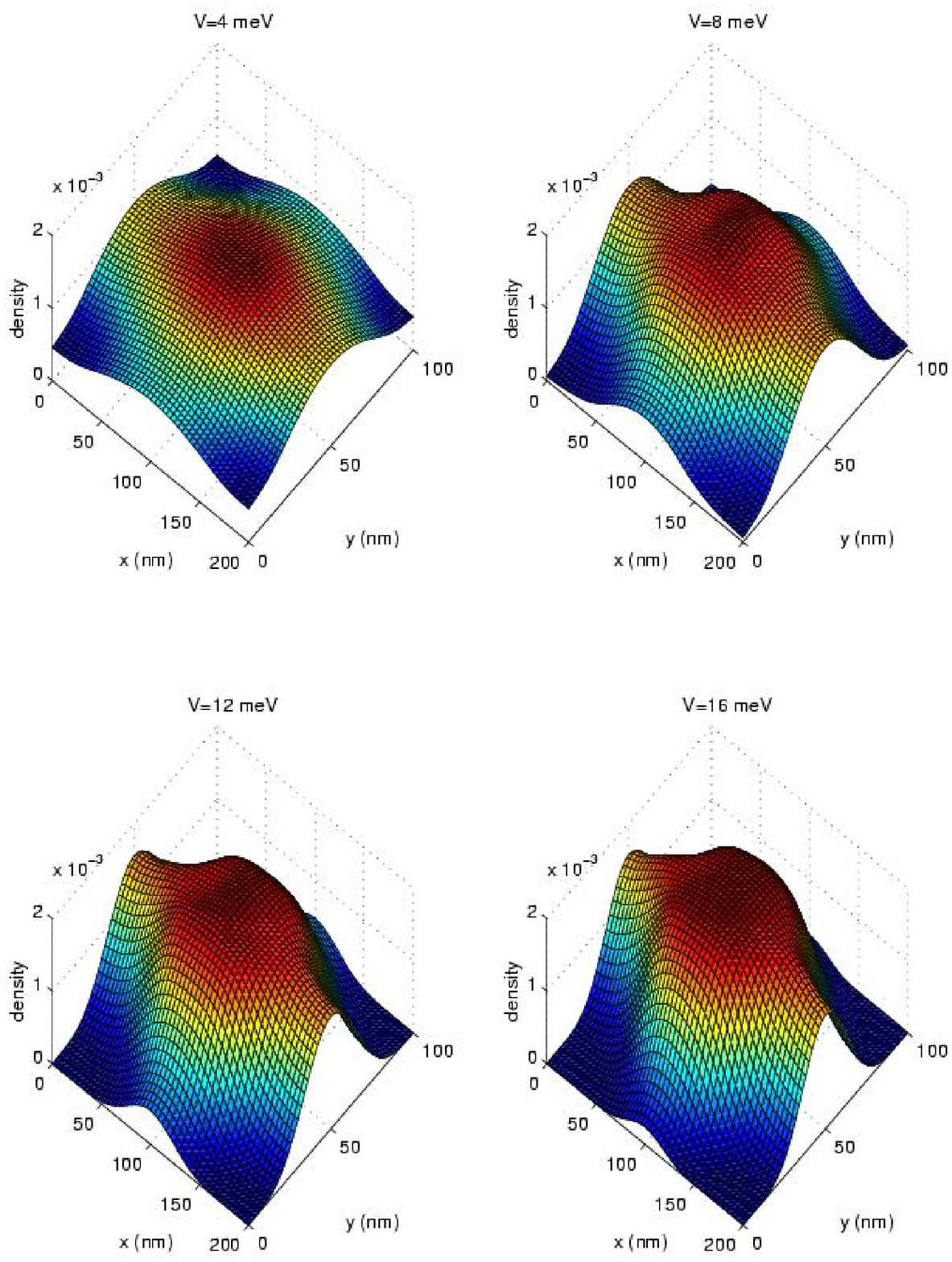}
\end{center}
\caption{The electron density distribution for the ground state of
         an {\it interacting} 2DEG in the simple periodic potential
         $V_{\mbox{\scriptsize per}}$ described by eq.\ (\ref{V_QAD}).
         The $x$ and $y$ axis are scaled differently here. 
         $B=1.654$ T, $T=1$ K, $N=20$.}
\label{RoVlda}
\end{figure}

This comparison opens the question of the role of the steepness of the
confining potential itself. To tackle that question we have redone
the calculations for the simple cosine potential $V_{\mbox{\scriptsize per}}$
defined by eq.\ (\ref{V_per}) with a variable strength $V_0$ but 
a constant number of electrons $N=20$. The results for the noninteracting
electrons is displayed in Fig.\ \ref{RoV0}. When $V_0$ is small the
overlapping of the electron density into neighboring cells is of the
same order of magnitude in both directions, but for strong modulation
$V_0$ the overlapping only takes place between the longer edges of the
cell, i.e.\ modulated wires are formed in the $y$ direction.
The curious fact is that absolutely the {\it contrary} happens for the interacting
system shown in Fig.\ \ref{RoVlda}, where the wires are formed in the
$x$ direction, which is the longer axis of the unit cell. Here several energy
bands are occupied in the case of 20 electrons in a unit cell. The
single-particle states with high energy are not well localized in the
minimum of the potential in the middle of the cell and thus the
wave functions of neighboring cells overlap where the distance is the
shortest, here in the direction of the short axis $L_y$.
With the interaction turned-on the repulsion in this direction is
stronger and the system forms wires in the direction with the shorter
interface with the neighboring cell. 

The softer confinement potential 
$V_{\mbox{\scriptsize per}}$ causes more drastic difference between
the interacting and the noninteracting electron system, than the more 
realistic dot confinement $V_{\mbox{\scriptsize QAD}}$, at least 
more realistic at the lattice lengths and number of electrons 
considered here.   

We have performed the calculations for quantum dots in arrays
with $L_x=L_y$ to confirm that no direction for 'wire formation' 
is preferred in that case, and when $L_y=1.5L_x$ we already 
have the wire formation in the preferred directions well developed. 
To test which parts of the dot interaction are important in influencing 
the shape we have repeated some of the calculations excluding the
exchange and correlation interaction. For the lattice lengths, 
the electron number, and the confinement selected here the exchange and 
correlation plays a minor role. There is a fine structure in the density
that depends on it, but the overall properties are caused by the
direct interaction as could be expected. 

\subsection{FIR absorption in the model of interacting dots}
Due to the large size of the mathematical set of basis functions used in the 
ground state calculation, in order to describe dots with several 
electrons and an array with not too short lattice lengths, we are not able to
perform a calculation of the FIR absorption for the system with the 
parameters used above. Instead, we can describe the electrons 
in the Hartree approximation (HA) without spin and in a smaller lattice
with $L_x=100$ nm and $L_y=150$ nm in a lower magnetic field strength
$B=1.10$ T. 

The FIR absorption is calculated in a self-consistent linear response
\cite{Gudmundsson96:5223R} exciting the system with an external 
electric field of the type
\begin{equation}
      {\bf E}_{ext}({\bf r},t)=-i{\mathcal E}_0\frac{{\bf k}+{\bf G}}
      {|{\bf k}+{\bf G}|}\exp{\left\{ i({\bf k}+{\bf G})\cdot{\bf r}
      -i\omega t\right\} } .
\label{Eext}
\end{equation} 
Here we do not restrict the
dispersion relation for the external field, $\omega({\bf k}+{\bf G})$,
to that of a free propagating electromagnetic
wave, but we allow for the more general situation in which the external
field is produced as in a near-field spectroscopy or in a
Raman scattering set-up.
The power absorption is found from the Joule heating of the self-consistent
electric field \cite{Dahl90:5763,Gudmundsson91:12098},
$-{\bf\nabla}\phi_{sc}$, with $\phi_{sc}=\phi_{ext}+\phi_{ind}$,
\begin{equation}
      P({\bf k}+{\bf G},\omega )=-\frac{\omega}{4\pi}\left[
      |{\bf k}+{\bf G}|\phi_{sc}({\bf k}+{\bf G},\omega )
      \phi_{ext}^*({\bf k}+{\bf G},\omega )\right] .
\label{Pom}
\end{equation}
The induced potential $\phi_{ind}$ is caused by the density 
variation $\delta n_s({\bf r})$ due to $\phi_{sc}$, 
which can then in turn be related to the
external potential by the dielectric tensor
\begin{equation}
      \sum_{{\bf G}'}\epsilon_{{\bf G},{\bf G}'}({\bf k},\omega )
      \phi_{sc}({\bf k}+{\bf G}',\omega )=\phi_{ext}({\bf k}+{\bf G},\omega ) .
\label{epsilon}
\end{equation}
The dielectric tensor, $\epsilon_{{\bf G},{\bf G}'}({\bf k},\omega )
=\delta_{{\bf G},{\bf G}'}-\frac{2\pi e^2}
{\kappa |{\bf k}+{\bf G}|}$  $\chi_{{\bf G},{\bf G}'}({\bf k},\omega )$,
is determined by the susceptibility of the electron system,
\begin{equation}
\begin{array}{r}
      \chi_{{\bf G},{\bf G}'}({\bf k},\omega )=\frac{1}
      {(2\pi L)^2}\int d{\bf \theta}
      \sum_{\alpha ,\alpha '}f_{\alpha ,\alpha '}
      ^{{\bf\theta},{\bf\theta}-{\bf k}L}(\hbar\omega )
      J_{\alpha ,\alpha '}^{{\bf\theta},{\bf\theta}-{\bf k}L-{\bf G}L}
      ({\bf k}+{\bf G})\\
      \left(J_{\alpha ,\alpha '}^{{\bf\theta},{\bf\theta}-{\bf k}L-{\bf G}L}
      ({\bf k}+{\bf G}')\right) ^* ,
\end{array}
\label{chi}
\end{equation}
where ${\bf k}$ is in the first Brillouin zone, $\tilde{{\bf k}}=(k_xL_x,k_yL_y)$
and $\tilde{{\bf G}}=(G_1L_x,G_2L_y)$, $\kappa$ is the dielectric
constant of the surrounding medium, 
${\bf\theta}=(\theta_1 ,\theta_2 )\in\{ [-\pi ,\pi]\times [-\pi ,\pi]\}$,
and
\begin{equation}
      f_{\alpha ,\alpha '}
      ^{{\bf\theta},{\bf\theta}'}(\hbar\omega )=
      \left\{ \frac{f^0(\varepsilon_{\alpha ,{\bf\theta}})-
      f^0(\varepsilon_{\alpha ' ,{\bf\theta}'})}
      {\hbar\omega+(\varepsilon_{\alpha ,  {\bf\theta}}-
                    \varepsilon_{\alpha ' ,{\bf\theta}'})+i\hbar\eta }
                                                                \right\},
\label{faa}
\end{equation}
in which $f^0$ is the equilibrium Fermi distribution, $\eta\rightarrow 0^+$,
and
\begin{equation}
           J_{\alpha ,\alpha '}^{{\bf\theta},{\bf\theta}'}({\bf k})=
           (\alpha '({\bf\theta}')|e^{-i{\bf k}\cdot{\bf r}}|\alpha (\theta)) .
\end{equation}
Special care must be taken with respect to the symmetry
of the wave functions corresponding to the Hartree states
$|\alpha{({\bf\theta}}))$ when translating them across the boundaries
of the quasi-Brillouin zones. 

\begin{figure}
\begin{center}
  \includegraphics[width=13.8cm]{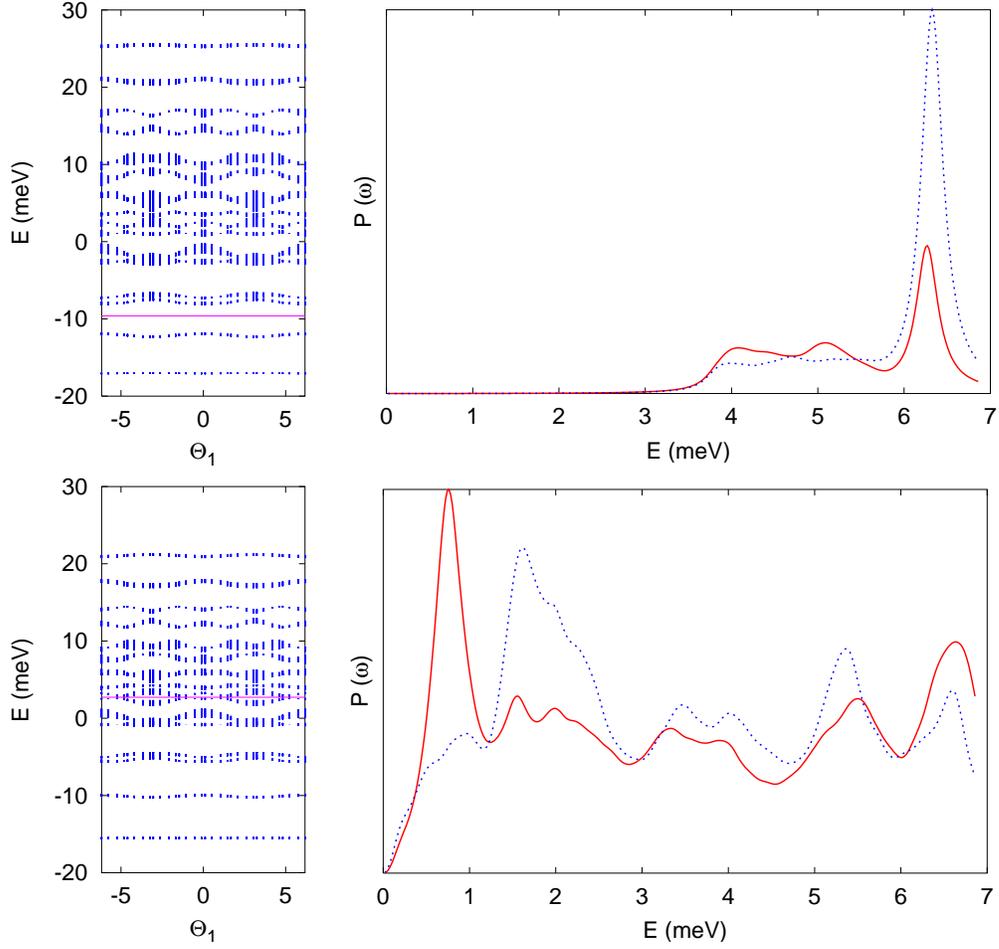}
\end{center}
\caption{The FIR absorption for two (upper panel) and seven electrons (lower panel)
         in the hartree approximation for spinless electrons. $k_xL_x=0.2$
         (solid curve), and $k_yL_y=0.2$ (dotted curve). The right panels 
         show the band structure projected on the $\Theta_1=k_xL_x$ axis in the
         Brillouin zone. The chemical potential $\mu$ is indicated by the
         horizontal solid line.}
\label{FIR}
\end{figure}

In Fig.\ \ref{FIR} we see the absorption for an array of electrons with
two electrons in each dot (upper right panel) and an array of seven electrons
(lower right panel). In the former case the dots are isolated, but in the latter
case their densities start to overlap in the $x$ and $y$ directions, almost
of the same amount. The structure of the corresponding energy bands is displayed
in the left panels of Fig.\ \ref{FIR}, showing that, indeed, when 7 electrons
occupy each dot the chemical potential $\mu$ is situated in the continuum 
states. We fix the polarization of the external field (\ref{Eext}) by giving
${\bf k}$ a small but finite value, ${\bf k}_iL_i=0.2$ with $i=x,y$, 
and ${\bf G}=0$ in accordance with FIR radiation.  
The FIR absorption of the isolated dots consists of two peaks, one sharp peak and 
the second broadened and lower. Since the confinement is not parabolic there are
higher order peaks at still higher energy that we exclude from our discussion
and figure here. At the higher electron number we still can locate the two
main peaks, now both split. At an energy below the collective dot modes, 
which are not dependent on the direction of ${\bf k}$, we find
intraband modes caused by transitions in the Landau band where
the chemical potential is located. These intraband modes depend on
the direction of ${\bf k}$ as the structure of the low lying continuum bands 
reflects the geometry of the dot array. They are generally not seen in
experimental spectra since for larger lattice lengths they are at a   
very low energy range that is not easily accessible.

As we can only consider a very limited system here, 
we have to be careful about generalizations, but if we analyze the gap
between the sets of peaks as a function of the electron number $N$ in 
the range between 2 and 12 electrons we can see a tendency to a 
similar evolution as has been reported in experiment \cite{Krahne01:xx} 
and is repeated in Fig.\ \ref{Fig_RK_elsqDisp}. 
This should only be considered as a very preliminary result and one has to
keep in mind that we perform our calculation at a finite magnetic 
field since the calculation is built on a basis set which has to increase
when the magnetic field decreases.

\subsection{Effects on magnetization in the ground state}
Recently, our attention has been drawn to measurements of the
magnetization of a homogeneous two-dimensional electron gas (2DEG) in heterostructures
\cite{Meinel99:819,Meinel01:121306}. There are efforts underway to
extend the experiments on magnetization also to modulated 2DEG's
and arrays of dots and antidots.
The magnetization is an equilibrium property of the ground state of the
electron system so we can calculate it for the system in which we have
studied the shape changes of the quantum dots as function of the
number of electrons $N$.
The total magnetization can be calculated according to the definition for
the orbital $M_o$ and the spin component of the magnetization $M_s$,
\cite{Desbois98:727,Gudmundsson00:4835}
\begin{equation}
      M_o+M_s=\frac{1}{2c{\mathcal A}}\int_{\mathcal A} d^2r \left( {\bf r}\times
      \langle {\bf J}({\bf r}) \rangle \right) \cdot\hat{{\bf n}}
      -\frac{g\mu_B}{{\mathcal A}}\int_{\mathcal A} d^2r \langle \sigma_z
      ({\bf r})\rangle \,,
\label{M_OS}
\end{equation}
where ${\mathcal A}$ is the total area of the system.
The equilibrium local current is evaluated as the quantum
thermal average of the current operator,
\begin{equation}
      \hat{{\bf J}}=-\frac{e}{2}\bigg(\,\hat{{\bf v}}|{\bf r}
      \rangle\langle{\bf r}|
      +|{\bf r}\rangle\langle{\bf r}|\hat{{\bf v}}\,\bigg) ,
\label{J_op}
\end{equation}
with the velocity operator
$\hat{\bf v}=[\hat{{\bf p}}+(e/c){\bf A}({\bf r})]/m^*$,
{\bf A} being the vector potential.

\begin{figure}
\begin{center}
  \includegraphics[width=8cm]{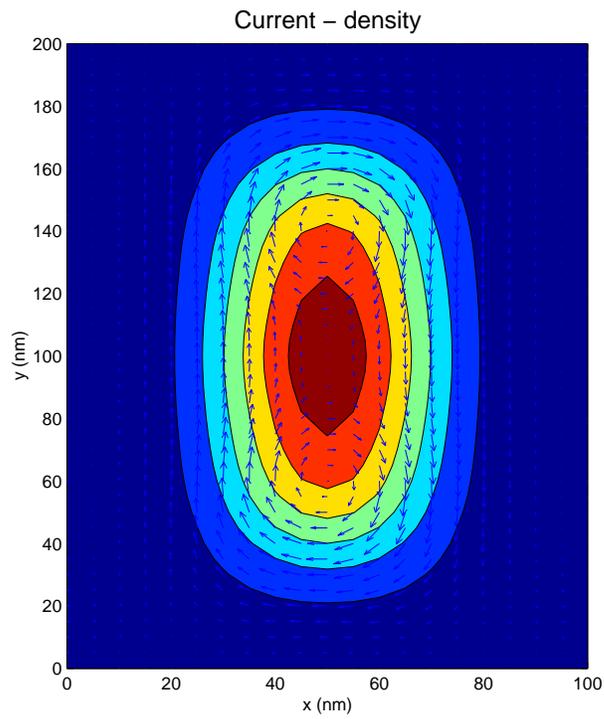}
\end{center}
\caption{The interacting electron and current density for 6 electrons
         in an elliptic quantum dot in the array described by eq.\
         (\ref{V_QAD}). $B=1.654$ T, $T=1$ K, $V_0=-16$ meV.}
\label{J_e2pq8_Ns6}
\end{figure}

\begin{figure}
\begin{center}
  \includegraphics[width=13.8cm]{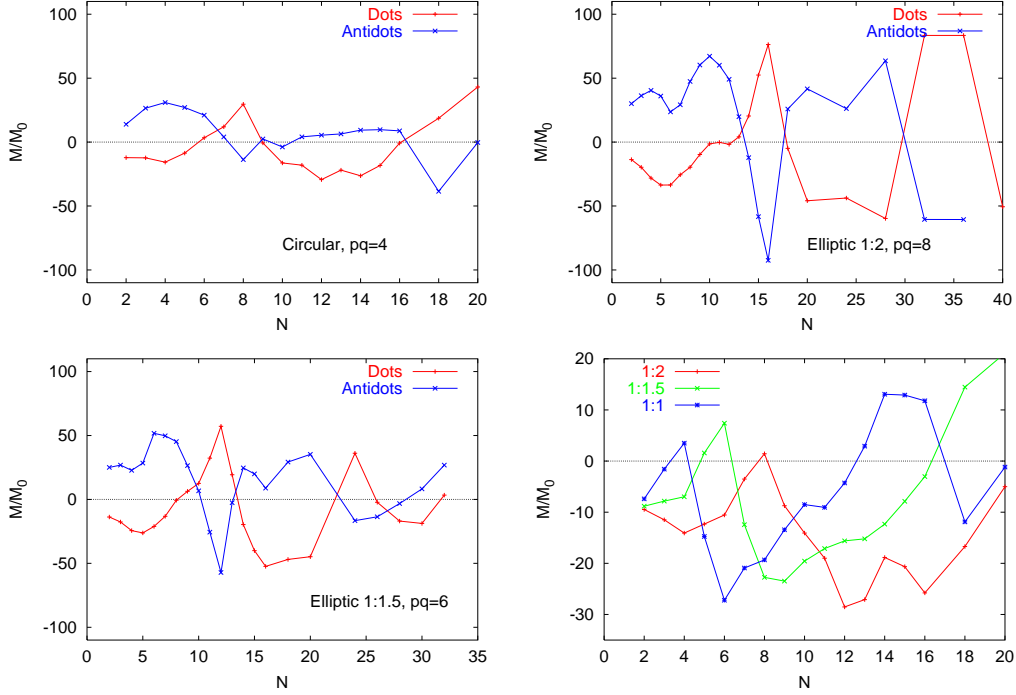}
\end{center}
\caption{The orbital magnetization $M_o$ as function of the number of electrons
         $N$ in a quantum dot or antidot array described by eq.\ (\ref{V_QAD}) 
         with aspect ratios 1:1 (upper left panel), 1:1.5 (upper right), 
         1:2 (lower left), a 2DEG confined by eq.\ (\ref{V_per}) for all three
         aspect ratios (lower right panel). $M_0=\mu_B^*/(L_xL_y)$,
         $B=1.654$ T, $T=1$ K,  $|V_0|=16$ meV. 
         }
\label{MAQD}
\end{figure}

\begin{figure}
\begin{center}
  \includegraphics[width=13.8cm]{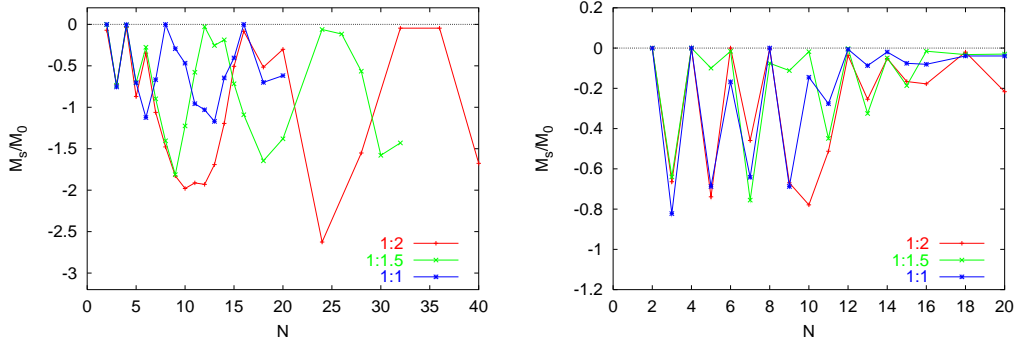}
\end{center}
\caption{The spin contribution to the magnetization $M_s$ as a function 
         of the number of electrons $N$ for a dot array described by
         eq.\ (\ref{V_QAD}) (left), and a 2DEG described by eq.\ (\ref{V_per}).
         $M_0=\mu_B^*/(L_xL_y)$, $B=1.654$ T, $T=1$ K,  $|V_0|=16$ meV.}
\label{MAQDs}
\end{figure}

A typical current density is shown in Fig.\ \ref{J_e2pq8_Ns6} superimposed on the
contours of the electron density for one elliptical quantum dot in an array of
dots. Even though the density has only one maximum two vortices are seen in 
the current density. Here again we have used the LSDA described above. 
The orbital magnetization of arrays of elliptical dots and antidots of different
aspect ratio is presented in Fig.\ \ref{MAQD} and for comparison the last
panel shows the magnetization for the electronic system confined by 
$V_{\mbox{\scriptsize per }}$ (\ref{V_per}). The magnetization for the antidot
array is almost simply the mirror image of the magnetization for the dot
array for the range of $N$ considered here, independent of whether the system
forms isolated dots or not. For low $N$ the $M_o$ develops peaks when $N$ equals twice
the number of flux quanta $pq$ through the unit cell, i.e.\ when only the lowest
Landau band is completely occupied and all other bands are empty. 
The spin contribution to the magnetization, $M_s$,
seen in the left panel of Fig.\ \ref{MAQDs}, reflects strong spin polarization  
as $N=pq$, when the first Landau band is half filled and the exchange energy is
thus enhanced. This enhancement of the exchange can also be recognized at higher
odd integer multiples of $pq$. 

The situation is a bit different for the electron system confined or
modulated by $V_{\mbox{\scriptsize per }}$ (\ref{V_per}). Here the 
splitting of the Landau bands into Hofstadter bands 
\cite{Hofstadter76:2239,Silberbauer92:7355,Gudmundsson95:16744} is stronger than 
the exchange enhancement of the spin splitting reflected by the fact
that the spin contribution to the magnetization $M_s$ in the right
panel of Fig.\ \ref{MAQDs} vanishes for even number of electrons in most
cases and no strong spin polarization is observed. This happens even 
when the iteration
process of the LSDA has been started with an artificial large $g$ factor that is 
later reduced to the natural value of 0.44 appropriate for GaAs.     
The last panel of Fig.\ \ref{MAQD} shows that the orbital magnetization
$M_o$ is also quite different for this system: First, its magnitude does
not increase as drastically with the size of the unit cell as for the
dots and antidots. Second, the Hofstadter splitting in the lowest Landau
band when it is half filled produces a clearer signature than the 
complete filling of the band. The difference in the magnetization for these
two systems has to be connected to their different geometry. At low $N$
the confinement $V_{\mbox{\scriptsize QAD}}$ produces simple dots or
antidots, the dots are isolated at first but start to overlap only after the
first Landau band has been fully occupied. On the other hand, the electron
system in $V_{\mbox{\scriptsize per}}$ forms connected regions for lower
$N$. At this moment we have not discovered any clear signs of the actual
geometry of the dots and antidots in the magnetization, 
and thus we can not distinguish the magnetization of circular or elliptic quantum dots.
In order to accomplish this in isolated dots with few electrons
we would need to be able to vary the magnetic field continuously
for a constant number of electrons \cite{Magnusdottir00:10229}.    

\section{Summary}
We have reported here on efforts to discern in experiments or predict by
model calculations the effects arrays can have on the FIR absorption of quantum
dots. There are indications that effects of the periodicity itself 
have been detected in measured FIR spectra, and even interaction between
neighboring quantum dots. Model calculations confirm that the effects of the
periodicity are well understood, but the effects of a direct interaction 
between the electron systems of different dots is very weak and subtle.
The direct interaction though seems to be detectable if it can 
influence the shape of the dots, since the FIR absorption is very dependent
on the symmetry of the electron system confined in them.  
   
%
%
\section*{Acknowledgments}
We gratefully acknowledge support from the German Science
Foundation DFG through SFB 508 ``Quanten-Materialien'', the
Graduiertenkolleg ``Nanostrukturierte Festk{\"o}rper'', the
Research Fund of the University of Iceland, and the Icelandic
Natural Science Council. We thankfully acknowledge the great
help of Birgir F.\ Erlendsson in parallelizing the execution
of the core regions of our programs.  
%
%
\bibliographystyle{apsrev}
\bibliography{mod_qd}

\newpage

\end{document}